\RequirePackage[2020-02-02]{latexrelease}
\documentclass[reprint,prb,showkeys,superscriptaddress,citeautoscript] {revtex4-2}

\usepackage{graphicx}
\usepackage{bm}
\usepackage{amsmath}
\usepackage{amsfonts}
\usepackage{amssymb}
\usepackage{array}
\RequirePackage{hyperref}
\hypersetup{%
  breaklinks = {true},
  citecolor = {blue},
  colorlinks = {true},
  linkcolor = {blue},
  pdfcreator = {\LaTeX\ and \flqq hyperref\frqq},
}
\hypersetup{colorlinks=true, urlcolor= blue, citecolor=blue, linkcolor= blue, bookmarks=true, bookmarksopen=false}

\def\k{{\bf k}}
\def\q{{\bf q}}

\begin{document}
		
		\title{First-principles prediction of superconductivity in MgB$_3$C$_3$}
		\author{Truong-Tho Pham}
		\affiliation{Laboratory of Magnetism and Magnetic Materials, Science and Technology Advanced Institute, Van Lang University, Ho Chi Minh City, Vietnam}
	\affiliation{Faculty of Applied Technology, School of Technology, Van Lang University, Ho Chi Minh City, Vietnam}		
		\author{Duc-Long Nguyen}
		\email{nguyenduclong@vlu.edu.vn}
		\affiliation{Laboratory of Applied Physics, Science and Technology Advanced Institute, Van Lang University, Ho Chi Minh City, Vietnam}		
		\affiliation{Faculty of Applied Technology, School of Technology, Van Lang University, Ho Chi Minh City, Vietnam}

\begin{abstract}
From first-principles density functional theory calculations, we propose hexagonal layered MgB$_3$C$_3$ as a potential phonon-mediated superconductor at 59 K, which is far higher than the superconductivity of MgB$_2$ ($\approx$ 39 K). The MgB$_3$C$_3$ is energetically and dynamically stable at ambient pressure in the \textit{P-62m} hexagonal structure with c/a $\approx$ 0.79 and forms in stacks of honeycomb B-C layers with Mg as a space filler. Band structure calculations indicate that the bands at the Fermi level derive mainly from B and C orbitals in which two $\sigma$- and two $\pi$ bands both contribute to the total density of state. The $\pi$ bands are found to be strongly couple with out-of-plane acoustic phonon mode, while the $\sigma$ bands coupled with the in-plane bond-stretching optical E$^{\prime} $ phonon modes produces a sizable superconductivity in MgB$_3$C$_3$.

\end{abstract}
	
\maketitle

	
\section{Introduction}
The discovery of the superconductor MgB$_2$ with a remarkably high superconducting transition temperature (T$_c$) of around 39 K has led to significant interest in studying the superconducting properties of graphite intercalation compounds at ambient pressure \cite{gao2020strong,singh2022high,haque2019first}. The mechanism behind the superconductivity of layered MgB$_2$-related compounds is thought to be the strong coupling between electrons in $p_{xy}$ orbitals ($\sigma$ bands) and the ultrafast phonon dynamics of the in-plane vibration of honeycomb layers, specifically the E$_{2g}$ stretching mode in MgB$_2$ \cite{novko2020ultrafast,liu2001beyond}. In contrast, the electron-phonon interaction is weaker in $p_z$ orbitals ($\pi$ bands) and contributes less to the electron-phonon coupling strength ($\lambda$) \cite{gao2020strong,quan2020li}. There have been many efforts in recent years to increase the T$_c$ of MgB$_2$ or to find new graphite intercalation compounds because these materials are still the best candidates for ambient pressure superconductors \cite{modak2021prediction,gao2015prediction,di2022high}. These efforts have focused on increasing the electronic density of states (DOS) at the Fermi level, and therefore the electron-phonon coupling, through chemical doping at the Mg or B sites, hydrogenation, or tensile strain. \cite{choi2009prediction,bekaert2019hydrogen,bekaert2017evolution}. One promising approach to finding new superconducting compounds involves replacing 50\% of the B atoms with C to form B-C honeycomb layers, resulting in compounds such as MgBC, LiBC, Mg$_2$B$_4$C$_2$, Li$_2$B$_3$C, etc. \cite{singh2022high,haque2019first,modak2021prediction,bazhirov2014electron}. Theoretical model calculations have suggested that the critical temperature (T$_c$) could reach 100 K in hole-doped Li$x$BC (x = 0.5) \cite{rosner2002prediction}. However, creating Li vacancies to dope the material with holes also produces local lattice distortion, which significantly changes the band structure, phonon vibrations, and electron-phonon interaction and ultimately destroys the superconductivity in Li$_{0.5}$BC. \cite{fogg2003libc}. While it has been proposed that boosting the critical temperature (T$_c$) by introducing Li deficiencies without causing lattice distortion is impractical \cite{haas2019color}, it is possible to select a unit cell containing a reduced number of Li (or Mg) atoms in practice. This has been a challenge for the past 20 years since the suggestion that metal deficiency could enhance the superconductivity of MB$_2$-like compounds (M: metal). In this work, we introduce a high-T$_c$ superconductor, MgB$_3$C$_3$, by reducing the number of Mg atoms by two-thirds (about 67\%) in MgB$_2$ while maintaining the periodic crystal structure without lattice distortion. Using a combination of first-principles calculations and isotropic Eliashberg theory, we show that the electron-phonon coupling between electrons from the $\sigma$ bands and the bond-stretching optical E$^{\prime}$ phonon modes at around 20.3 THz (677 cm$^{-1}$, 83 meV) is strengthened in MgB$_3$C$_3$. The interaction of electrons in the $\pi$ bands with the out-of-plane acoustic modes of the B-C lattice also contributes to an increase in $\lambda$ up to 1.1, resulting in a maximum T$_c$ of 59 K using the same parameters calculated for MgB$_2$ (38 K).

\begin{figure*}
\includegraphics[width=16cm]{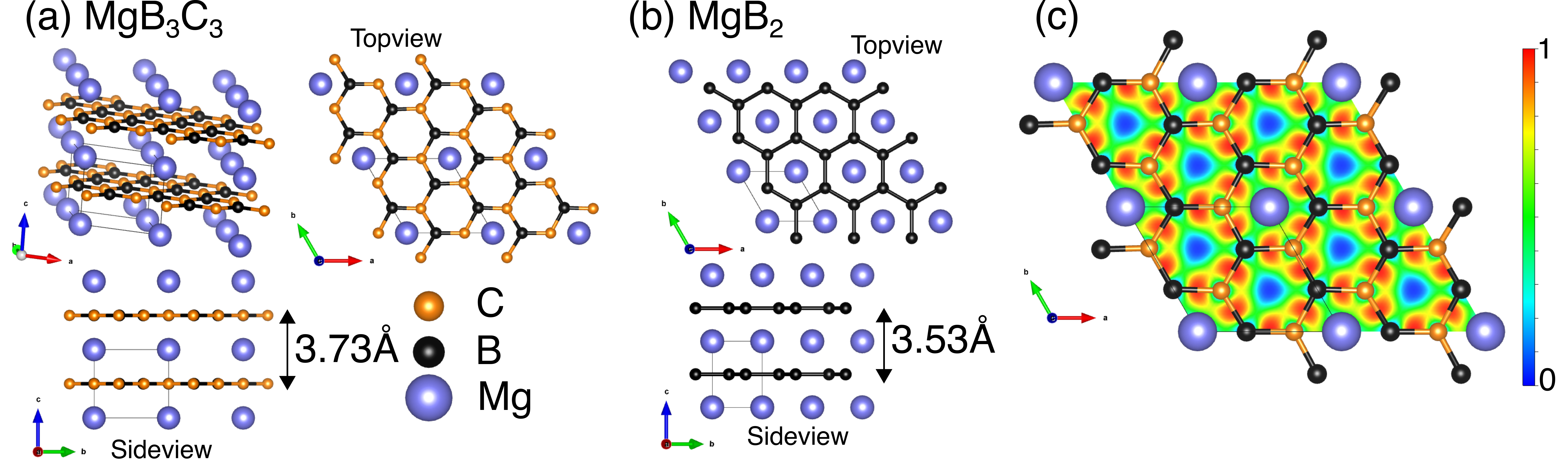}
  \caption{(Color online) (a) Top and side view of MB$_3$C$_3$, (b) top and side view of MgB$_2$. The orange balls represent C atoms, black and purple blue balls are B and Mg atoms, The thin line denotes the unit cell. (c) Electron localization function of MB$_3$C$_3$ on (001) plane. The color bar represents the value of ELF from 0 to 1.}\label{fig:crystal_structure}
\end{figure*}

\section{Computational methods}
Using density functional theory \citep{DFT01_hohenberg1964inhomogeneous,dft1965_kohn1965self} as implemented in the Quantum ESPRESSO package \citep{QUANTUMESPRESO03_giannozzi2020quantum,QUANTUMESPRESSO02_giannozzi2017advanced}, we conducted geometry optimizations of both atomic positions and lattice parameters, as well as electronic structure calculations including band structures, densities of states, and Bader charges\cite{yu2011accurate}. The generalized gradient approximation (GGA) in the form of Perdew-Burke-Ernzerhof(PBE) exchange and correlation functional \cite{PBE_perdew1996generalized} and the standard solid-state pseudopotentials library \cite{SSSP01_prandini2018precision} for pseudopotentials were used. The plane-wave basis set energy cutoff and charge density cutoff were set to 45 Ry and 360 Ry, respectively. The k-point mesh used was 18$\times$18$\times$18 with the $\Gamma$-centered Monkhorst-Pack scheme \citep{KPOINTS_monkhorst1976special}. The phonon dispersion and the electron-phonon coupling constants are calculated based on density functional perturbation theory (DFPT) \cite{baroni2001phonons} and Eliashberg equations \cite{eliashberg1960interactions} with a 6$\times$6$\times$6 \textbf{\textit{q}}-point grid. We employed linear interpolation to interpolate the electron-phonon coupling matrix to a denser grid of 36$\times$36$\times$36 in order to obtain a converged result\cite{wierzbowska2005origins}. To calculate the critical temperature, T$_c$, of MgB$_3$C$_3$, we also used the OptB88-vdW exchange-correlation functionals \cite{thonhauser2015spin}with van der Waals (vdW) correction. This method has been commonly used for the structural optimization of van der Waals layered materials and has been found to have better agreement with experimental results. An evolutionary algorithm was used to find the crystal structure of MgB$_3$C$_3$ as implemented in the Cryspy package \cite{yamashita2022hybrid}. This algorithm works by starting with a population of structures and evolving them through a process of "survival of the fittest," in which structures with lower energy become the parents of a new generation of structures \cite{zhou2014semimetallic}. We used tournament selection and roulette selection to select the parent structures from the population of surviving parents. To produce offspring from the parents, we used evolutionary operations such as crossover, permutation, and strain, which were set to 10, 3, and 5, respectively. The new generation of structures included not only those produced by these evolutionary operations, but also random structures and structures chosen through elite selection throughout 30 generations. We included two new random structures and two elite selection structures in each generation. The result of crystal structure prediction is shown in Fig. S1 Supplementation information.
\section{Main results and discussion}
The crystal structure of MgB$_3$C$_3$ is shown in Fig.\ref{fig:crystal_structure}. It can be seen that MgB$_3$C$_3$ belongs to a class of graphite intercalation compounds, with the B-C lattice forming stacks of honeycomb layers and Mg as a spacer. The Mg atoms are located at the \textit{1b} (0,0,0.5) Wyckoff position, while the B and C atoms occupy the \textit{3f} (\textit{x},0,0) position, with \textit{x} values of 0.65846 and 0.32597 for B and C, respectively. The formation energy of MgB$_3$C$_3$ is calculated to be 0.0038 eV/atom, which is a positive value but small enough to be potentially achievable in experiments. In contrast to the crystal structure of MgB$_2$ (Figure \ref{fig:crystal_structure}b), the number of Mg atoms is critically reduced to one-third (about 33\%), corresponding to a deficiency of two-thirds (about 67\%) of the Mg atoms in MgB$_2$ (or MgBC)\cite{haque2019first}. It is worth noting that the Mg deficiencies in MgB$_3$C$_3$ are much higher than the 50\% Li deficiencies in LiBC\cite{rosner2002prediction}. Despite this, the lattice structure of MgB$_3$C$_3$ maintains a non-distorted configuration with uniform honeycomb B-C layers, as shown in Figure \ref{fig:crystal_structure}(a). The bond length between B and C is around 1.55 \AA\ and 1.56 \AA, which is smaller than the 1.77 \AA\ bond length in the B-B honeycomb of MgB$_2$. This nonuniformity in bond length appears to be a characteristic feature of B-C layers \cite{gao2015prediction}, which would increase a degree of freedom in the phonon vibrations of lattice. In addition, a shorter bond length also leads to better orbital overlap between B and C atoms, resulting in stronger $\pi$ and $\sigma$ bands that couple more strongly with phonon vibrations. High electron localization function (ELF) values above 0.9 (see Fig. \ref{fig:crystal_structure}(c)) near the center of the bonds suggest the presence of strong covalent bonding throughout the B-C ring in MgB$_3$C$_3$. The spacing between two adjacent honeycomb B-C layers is about 3.73 \AA, which is slightly higher than that of 3.5 \AA\ for Mg$_2$B$_4$C$_2$, 3.53 \AA\ for MgB$_2$, and 3.38 \AA\ for Li$_4$B$_5$C$_3$ \cite{singh2022high,bazhirov2014electron}. The unit cell of MgB$_3$C$_3$ with a hexagonal structure (space group \textit{P-62m}) is shown as thin lines in \ref{fig:crystal_structure}(a), with lattice constants of a = 4.69 \AA\ and c = 3.73 \AA. Most MgB$_2$-like superconducting compounds have a hexagonal unit cell with a c/a ratio greater than 1, so the c/a ratio of approximately 0.79 in MgB$_3$C$_3$ is unusual. Therefore, the electronic band structure is expected to be significantly different from previous reports.

\begin{figure}
  \includegraphics[width=9cm]{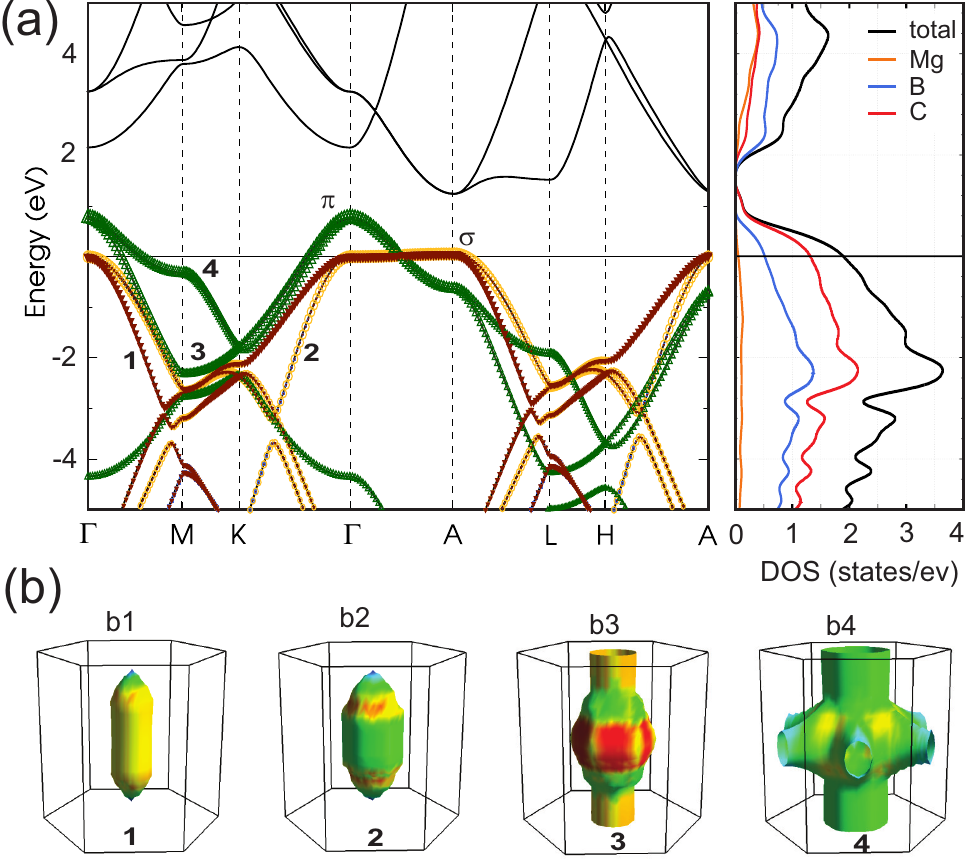}
  \caption{(Color online) . Electronic structure of MgB$_3$C$_3$. (a) Band structure. The $\sigma$- and $\pi$ bands crossing the Fermi level are indicated (b1)-(b4) Fermi surfaces corresponding to the four bands labeled in (a).}\label{fig:band_structure}
\end{figure}
The electronic band structure, electronic densities of states, and Fermi surfaces of MgB$_3$C$_3$ are thus shown in Fig.\ref{fig:band_structure}. The energy windows of the electronic band structure of MgB$_3$C$_3$ within 5 eV of the Fermi level are illustrated here as the most significant feature of the material. The high-energy portion of the valence band in MgB$_3$C$_3$ consists of two $\pi$ bands and two $\sigma$ bands that cross the Fermi level and create a small indirect gap of 0.487 eV between the valence and conduction bands. This type of gap between the $\sigma$-bonding and $\sigma^*$-antibonding bands is not commonly observed in MgB$_2$-related compounds, with the exception of LiB$_2$C$_2$ trilayer \cite{gao2020strong}. The $\pi$ bands in MgB$_3$C$_3$ exhibit a nearly flat band along the $\Gamma$-A direction of the Brillouin zone, similar to MgB$_2$, but these bands are almost completely filled, with their maximum energies just above the Fermi level by about 0.05 eV, and they incline from the $\Gamma$ to A points. These bands give rise to a small hole pocket and a spindle-shaped hole Fermi surfaces, as shown in \ref{fig:band_structure}(b1) and (b2). The two $\sigma$ bands of MgB$_3$C$_3$, which cross the Fermi level, have a relative energy that is similar to that of the two $\pi$ bands throughout the Brillouin zone.The Fermi surface of the two $\sigma$ bands is shown in \ref{fig:band_structure}(b3) and (b4). It has been previously shown that the lowering of the $\sigma$ bands relative to the $\pi$ bands leads to a $\pi$ $\rightarrow$ $\sigma$ charge transfer and $\pi$-band hole doping effect, which is believed to be the driving force behind the superconductivity in MgB$_2$ \cite{an2001superconductivity,kurmaev2002electronic}. We would like to emphasize that the lowering of the $\sigma$ bands is due to the transfer of charge from valence electrons in Mg 3s to B 2$pz$ orbitals, which causes a slight incline in the flat band along the $\Gamma$-A direction \cite{kim2018electronic,ravindran2001detailed}. In MgB$_3$C$_3$, the $\sigma$- and $\pi$ bands have comparable energy levels with multiple crossover points, so the transfer of charge from Mg to the B-C lattices and the transition from $\pi$ to $\sigma$ bands may also be beneficial for structural stability and superconductivity. The Bader charge analysis shows that Mg donates an average of 1.6E$-$e to the B-C ring. It is worth noting that the $\sigma$ bands also contribute to the lower energy bands (below -2 eV). The $\sigma$ (or $\pi$) bands of the valence band have a higher occupancy in the 2$p_z$ orbital (or pxy orbital) of C atoms compared to that of B atoms, as seen in the density of states profile. However, the occupancy of the 2$pz$ orbital of C atoms is smaller in the conduction band. On the other hand, the weighted occupancy of Mg s orbitals almost disappears near the Fermi level, but it is present at higher and lower energy levels (see Fig. S2 of Supplementation information); This indicates that there is a charge transfer from Mg to the B-C layers \cite{modak2021prediction}. The substantial weight of Mg atoms in the density of states (DOS) profile is mainly due to the 2p orbitals. The total DOS at the Fermi level is around 1.9 states/eV, mostly from the contribution of the $\pi$- and $\sigma$ bands of C (68\%) and B (26\%). In contrast, Mg atoms only contribute 6\% to the DOS. It is worth noting that MgB$_3$C$_3$ has a higher DOS at the Fermi level compared to MgB$_2$ and related compounds \cite{gao2020strong,quan2020li,bekaert2019hydrogen}. 

\begin{figure}
  \includegraphics[width=9cm]{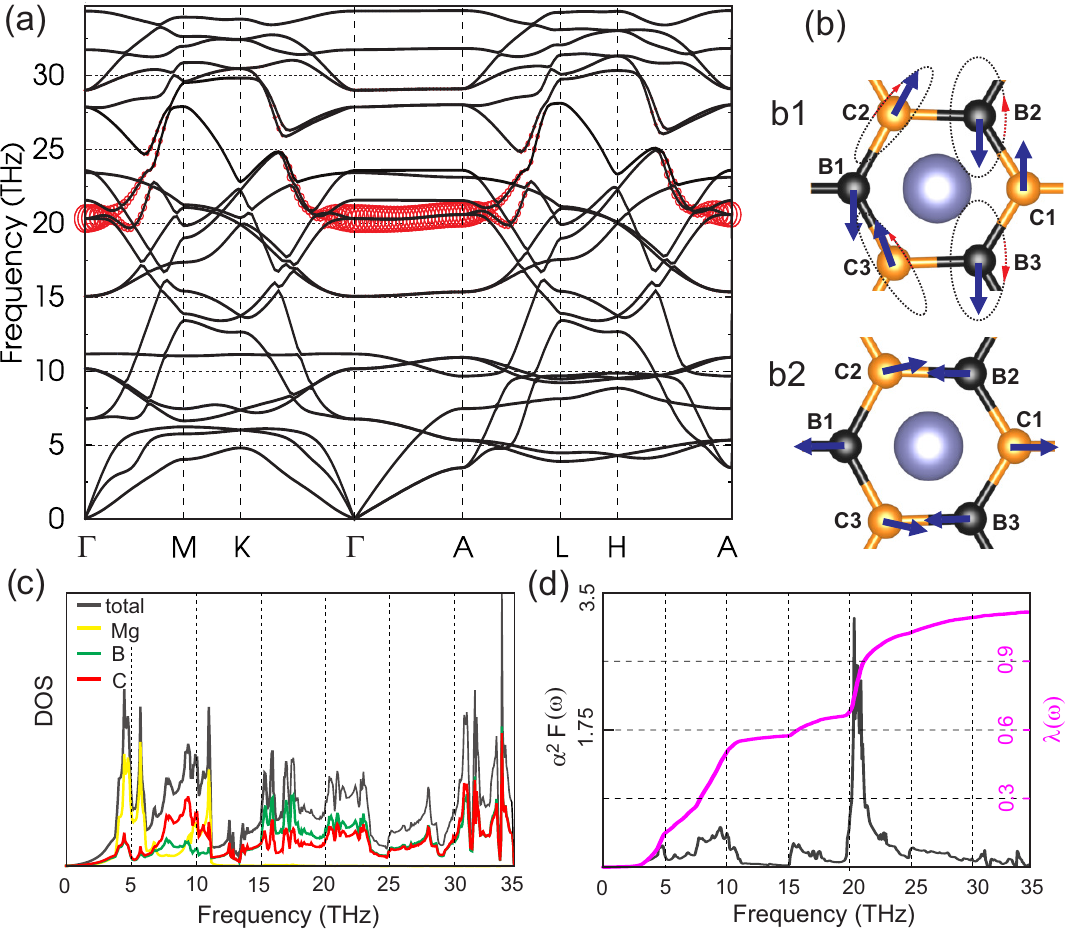}
  \caption{(Color online) (a) Phonon dispersion curves in MgB$_3$C$_3$. The size of red circles represents the strength of the electron-phonon coupling. (b) The twofold phonon vibration for double degenerate E’ mode along the $\Gamma$-A direction (20.3 THz) having strong coupling with electrons. (c) phonon density of states including atom-resolved contributions. (d) Eliashberg spectral functions $\alpha^2F(\omega)$) and electron-phonon coupling strengths $\lambda(\omega)$}.\label{fig:electron_phonon}
\end{figure}

Next, we will discuss the lattice dynamics and electron-phonon coupling (EPC) of MgB$_3$C$_3$. Using density functional perturbation theory (DFPT) \cite{baroni2001phonons}, we can calculate the phonon frequencies by diagonalizing the dynamical matrix, and the electron-phonon coefficients $g$ can be obtained from the first-order derivative of the self-consistent Kohn-Sham (KS) potential:
\begin{equation}
g_{\q\nu}(\k,i,j) =\left({\hbar\over 2M\omega_{\q\nu}}\right)^{1/2}
\langle\psi_{i,\k}| {dV_{SCF}\over d {\hat u}_{\q\nu} }\cdot
                   \hat \epsilon_{\q\nu}|\psi_{j,\k+\q}\rangle.
\end{equation}
where $M$ represents the atomic mass, ${\bf q}$ and ${\bf k}$ are wave vectors, and $ij$ and $\nu$ refer to indices of electronic energy bands and phonon modes, respectively.
The phonon linewidth $\gamma_{\q\nu}$ is defined by
\begin{equation}
\begin{aligned}
\gamma_{\q\nu} = 2\pi\omega_{\q\nu} \sum_{ij}
                \int {d^3k\over \Omega_{BZ}}  |g_{\q\nu}(\k,i,j)|^2 \\
                    \delta(e_{\q,i} - e_F)  \delta(e_{\k+\q,j} - e_F), 
\end{aligned}
\end{equation}
The EPC constant, $\lambda_{\q\nu}$, for mode $\nu$ at wavevector $\q$ is defined as
\begin{equation}
\lambda_{\q\nu} ={\gamma_{\q\nu} \over \pi\hbar N(e_F)\omega^2_{\q\nu}}
\end{equation}

\begin{table*}[!htb]
\centering \caption{The superconducting parameters of fully relaxed MgB$_3$C$_3$ and MgB$2$, including  density of states (DOS) at the Fermi energy, and the superconducting transition temperature ($T_c$), were calculated using different functionals. The units for DOS are states/spin/Rydberg/unit cell, and the units for $\omega_{ln}$ and $T_c$ are K. }\label{tab:superconductivity}
  \begin{tabular*}{0.8\textwidth}{@{\extracolsep{\fill}}ccccccccc}
  \hline\hline
Compound& Functional &$\lambda$ & DOS(E$_F$)& $\omega_{ln}$(K)& T$_c^{\mu^*0.1}$(K)& T$_c^{\mu^*0.13}$(K) & T$_c^{\mu^*0.05}$(K)&\\\hline
MgB$_3$C$_3$ &PBE& 1.09 & 12.67&572.9 & 45.3 & 39.6 & 55.1\\\hline
MgB$_3$C$_3$ &OptB88-vdW&1.16&12.76&573.6&49.3&43.5&59.2\\\hline
MgB$_2$ &PBE&0.67& 4.71&697.8&21.4&16.0&31.6\\\hline
MgB$_2$ &OptB88-vdW&0.74&4.73&682.9&27.4&21.6&38.1\\\hline
\end{tabular*}
\end{table*}
Fig. \ref{fig:electron_phonon} shows the phonon dispersions, phonon DOS, Eliashberg spectral functions $\alpha^2F(\omega)$, electron-phonon coupling strengths $\lambda(\omega)$, and the representative phonon vibrations for double degenerate E$^{\prime} $ mode. The phonon dispersions in Fig. \ref{fig:electron_phonon}(a) show no imaginary phonon modes, suggesting that MgB$_3$C$_3$ is dynamically stable at ambient pressure. Five low-lying optical modes exist at the zone center $\Gamma$, two of which are degenerate modes at frequencies of 6.76 and 10.46, and one at 11.2 THz. These modes involve the in-plane and out-of-plane vibrations with governing contribution of Mg and C atoms to the phonon DOS, as seen in Fig. \ref{fig:electron_phonon}(c). We have observed that the double degenerate mode E$^{\prime\prime}$ at 10.46 THz corresponds to a specific out-of-plane vibration of the C atoms. The other optical modes located at 15.1, 20.3, 21.6, 23.4, 23.6, 27.9, 29.0, 31.7, and 34.3 THz at the zone center $\Gamma$ are solely attributed to vibrations of the B-C lattice. All of these optical modes are found to be quite dispersive, with the exception of those along the $\Gamma$-A direction, which reflects the layered structure of MgB$_3$C$_3$. Among these phonon modes, the double degenerate mode at 20.3 THz (677 cm$^{-1}$, 83 meV) is of particular interest when it has a sizable coupling with electrons at the Fermi level (denoted by red circles in Fig. \ref{fig:electron_phonon}(a)), which is similar to the E$_{2g}$ mode in MgB$_2$ \cite{yildirim2001giant}. This mode involves the degenerate in-plane bond-stretching E$^{\prime}$ modes along the $\Gamma$-A direction, which are shown in figure \ref{fig:electron_phonon}(b1) and (b2). The arrows and ellipse in the figure indicate the movement of the atoms: the lower frequency mode involves the rotation of an ellipse and the backward and forward movement of the B and C atoms, while the higher frequency mode only involves the back and forth movement. These E$^{\prime}$ modes are very anharmonic which evident of a large phonon linewidth and can achieve a significantly large effective population due to their ultrafast phonon dynamics \cite{novko2020ultrafast}. We observed that the vibration frequency of the in-plane bond-stretching E$^{\prime}$ mode in MgB$_3$C$_3$ is much higher than in MgB$_2$ (74.5 meV), but similar to that in monolayer LiBC (86.5 meV) \cite{modak2021prediction} and LiB$_2$C$_2$ trilayer film (87.7 meV)\cite{gao2020strong}. We want to emphasize that, although the in-plane bond-stretching E' mode does not contribute significantly to the phonon density of states compared to other modes, its strong coupling with electrons near the Fermi level produces a very intense peak in the Eliashberg spectral function (defined as $\alpha^2F(\omega) = {1\over 2\pi N(e_F)}\sum_{\q\nu} 
                    \delta(\omega-\omega_{\q\nu})
                    {\gamma_{\q\nu}\over\hbar\omega_{\q\nu}}.$), as seen in \ref{fig:electron_phonon}(d). Three acoustic modes make a major contribution to the Eliashberg spectral function in the low frequency range (less than 10.8 THz), accounting for nearly 48\% of the total electron-phonon coupling $\lambda = \sum_{\q\nu} \lambda_{\q\nu} = 
2 \int {\alpha^2F(\omega) \over \omega} d\omega.$. A sizable contribution of low-frequency phonon modes to the $\lambda(\omega)$ is in good agreement with previous reports \cite{gao2020strong,modak2021prediction,novko2020ultrafast}. The acoustic phonon-mediated superconductivity also reports previously in MBC (M = Ba, Sr, etc.) \cite{haque2019first}. In MgB$_3$C$_3$, the Mg atoms have a significant presence in the phonon density of states at low frequencies, but their in-plane vibrations have less impact on $\alpha^2F(\omega)$ and $\lambda(\omega)$ due to weak orbital overlap between Mg atoms. Instead, the low-frequency spectra weight in $\alpha^2F(\omega)$ and $\lambda(\omega)$ is mainly due to the out-of-plane displacement of B and C atoms, indicating important coupling between the $\pi$ bands and low-frequency modes. This is not the case in MgB$2$, where the high-frequency (87.1 meV) out-of-plane vibration of boron honeycomb layers (B${1g}$ mode) has less impact on $\alpha^2F(\omega)$ and $\lambda(\omega)$ \cite{margine2013anisotropic}. It has been observed that the coupling between electron $\pi$ bands and acoustic phonons is enhanced in honeycomb B-C lattices, as well as between $\pi$ bands and the out-of-plane vibration of B-C lattices in the optical branch (specifically, the modes at 15.1 and 21.6 THz of the E$^{\prime}$ point). This contributes to $\alpha^2F(\omega)$ and $\lambda(\omega)$, as shown in Figure 3(d). The out-of-plane vibration of B atoms in the frequency range of 15-20 THz contributes approximately 12\% to $\lambda(\omega)$, while other optical modes above 20 THz contribute about 40\%. The total electron-phonon coupling constant is approximately 1.1, which is about 25\% larger than that of MgB$_2$.

We then obtained the superconducting parameters of MgB$_3$C$_3$ as summarized in Tab.\ref{tab:superconductivity}. For comparison, we also calculated the same parameters for MgB$_2$ using a similar DFPT setup. It is known that using different exchange correlation functionals can lead to different results, so we also used the vanderWall correction functional OptB88-vdW to estimate the Allen-Dynes-modified McMillan formula $T_c = {\omega_{log}\over 1.2} \mbox{exp} \left [ 
         {-1.04(1+\lambda)\over \lambda(1-0.62\mu^*)-\mu^*}\right ]$ for both cases. This functional has been shown to provide better agreement with experimental results for MgB$_2$, even without solving the anisotropic Eliashberg equations \cite{an2021superconductivity}. Although McMillan formula relies on the semi-empirical parameter of Coulomb repulsion ($\mu^*$), the predictive power of this approach remains significant. We obtained a superconducting transition temperature (T$_c$) of 45.3 K using the commonly used value of $\mu^*$=0.1, which is double the value obtained for MgB$_2$ using the same parameter. It is worth noting that the calculated value of $\lambda$ for MgB$_2$ is in good agreement with previous theoretical studies \cite{kawamura2014improved,an2021superconductivity,choi2002first} and experiments \cite{bouquet2001specific}.
The OptB88-vdW functional resulted in a higher value of $\lambda$ for both MgB$_2$ and MgB$_3$C$_3$. As a result, the critical temperature (T$_c$) of MgB$_2$ was calculated to be 27.4 K with a value of $\mu^*$ = 0.1, but to match the experimental value, a value of $\mu^*$ = 0.05 is needed. On the other hand, the very same value resulted in a T$_c$ of 59.2 K for MgB$_3$C$_3$. Based on this numerical analysis, it can be concluded that MgB$_3$C$_3$ may be a superior superconductor compared to MgB$_2$ due to its unique crystal structure, electronic structure, and the electron-phonon coupling nature.
\section{Conclusion}
To summarize, we have introduced a high-temperature superconductor, MgB$_3$C$_3$, by reducing the number of Mg atoms by two-thirds (67\%) while maintaining the periodic crystal structure without lattice distortion, using crystal structure prediction and first principles calculations. We have shown the electron-phonon coupling between electrons from the $\sigma$ bands and the bond-stretching optical E' phonon modes at 20.3 THz is strengthened in MgB$_3$C$_3$. The interaction of electrons in the $\pi$ bands with the out-of-plane acoustic modes of the B-C lattice also contributes to an increase in $\lambda$ up to 1.1, resulting in a maximum critical temperature of 59 K using the same parameters calculated for MgB$_2$ (38 K). 

\section*{Acknowledgment}
The authors would like to thank for the support from Van Lang University. We would also like to express our appreciation to Prof. Ching-Ming Wei at the Institute of Atomic and Molecular Sciences at Academia Sinica in Taipei, Taiwan for generously providing us with the computational resources needed to complete this study.	
\section*{Declaration of Interests}
The authors declare that they have no conflict of interest.
	

\begin{thebibliography}{40}%
\makeatletter
\providecommand \@ifxundefined [1]{%
 \@ifx{#1\undefined}
}%
\providecommand \@ifnum [1]{%
 \ifnum #1\expandafter \@firstoftwo
 \else \expandafter \@secondoftwo
 \fi
}%
\providecommand \@ifx [1]{%
 \ifx #1\expandafter \@firstoftwo
 \else \expandafter \@secondoftwo
 \fi
}%
\providecommand \natexlab [1]{#1}%
\providecommand \enquote  [1]{``#1''}%
\providecommand \bibnamefont  [1]{#1}%
\providecommand \bibfnamefont [1]{#1}%
\providecommand \citenamefont [1]{#1}%
\providecommand \href@noop [0]{\@secondoftwo}%
\providecommand \href [0]{\begingroup \@sanitize@url \@href}%
\providecommand \@href[1]{\@@startlink{#1}\@@href}%
\providecommand \@@href[1]{\endgroup#1\@@endlink}%
\providecommand \@sanitize@url [0]{\catcode `\\12\catcode `\$12\catcode
  `\&12\catcode `\#12\catcode `\^12\catcode `\_12\catcode `\%12\relax}%
\providecommand \@@startlink[1]{}%
\providecommand \@@endlink[0]{}%
\providecommand \url  [0]{\begingroup\@sanitize@url \@url }%
\providecommand \@url [1]{\endgroup\@href {#1}{\urlprefix }}%
\providecommand \urlprefix  [0]{URL }%
\providecommand \Eprint [0]{\href }%
\providecommand \doibase [0]{https://doi.org/}%
\providecommand \selectlanguage [0]{\@gobble}%
\providecommand \bibinfo  [0]{\@secondoftwo}%
\providecommand \bibfield  [0]{\@secondoftwo}%
\providecommand \translation [1]{[#1]}%
\providecommand \BibitemOpen [0]{}%
\providecommand \bibitemStop [0]{}%
\providecommand \bibitemNoStop [0]{.\EOS\space}%
\providecommand \EOS [0]{\spacefactor3000\relax}%
\providecommand \BibitemShut  [1]{\csname bibitem#1\endcsname}%
\let\auto@bib@innerbib\@empty
\bibitem [{\citenamefont {Gao}\ \emph {et~al.}(2020)\citenamefont {Gao},
  \citenamefont {Yan}, \citenamefont {Lu},\ and\ \citenamefont
  {Xiang}}]{gao2020strong}%
  \BibitemOpen
  \bibfield  {author} {\bibinfo {author} {\bibfnamefont {M.}~\bibnamefont
  {Gao}}, \bibinfo {author} {\bibfnamefont {X.-W.}\ \bibnamefont {Yan}},
  \bibinfo {author} {\bibfnamefont {Z.-Y.}\ \bibnamefont {Lu}},\ and\ \bibinfo
  {author} {\bibfnamefont {T.}~\bibnamefont {Xiang}},\ }\href@noop {}
  {\bibfield  {journal} {\bibinfo  {journal} {Physical Review B}\ }\textbf
  {\bibinfo {volume} {101}},\ \bibinfo {pages} {094501} (\bibinfo {year}
  {2020})}\BibitemShut {NoStop}%
\bibitem [{\citenamefont {Singh}\ \emph {et~al.}(2022)\citenamefont {Singh},
  \citenamefont {Romero}, \citenamefont {Mella}, \citenamefont {Eremeev},
  \citenamefont {Mu{\~n}oz}, \citenamefont {Alexandrova}, \citenamefont {Rabe},
  \citenamefont {Vanderbilt},\ and\ \citenamefont {Mu{\~n}oz}}]{singh2022high}%
  \BibitemOpen
  \bibfield  {author} {\bibinfo {author} {\bibfnamefont {S.}~\bibnamefont
  {Singh}}, \bibinfo {author} {\bibfnamefont {A.~H.}\ \bibnamefont {Romero}},
  \bibinfo {author} {\bibfnamefont {J.~D.}\ \bibnamefont {Mella}}, \bibinfo
  {author} {\bibfnamefont {V.}~\bibnamefont {Eremeev}}, \bibinfo {author}
  {\bibfnamefont {E.}~\bibnamefont {Mu{\~n}oz}}, \bibinfo {author}
  {\bibfnamefont {A.~N.}\ \bibnamefont {Alexandrova}}, \bibinfo {author}
  {\bibfnamefont {K.~M.}\ \bibnamefont {Rabe}}, \bibinfo {author}
  {\bibfnamefont {D.}~\bibnamefont {Vanderbilt}},\ and\ \bibinfo {author}
  {\bibfnamefont {F.}~\bibnamefont {Mu{\~n}oz}},\ }\href@noop {} {\bibfield
  {journal} {\bibinfo  {journal} {npj Quantum Materials}\ }\textbf {\bibinfo
  {volume} {7}},\ \bibinfo {pages} {1} (\bibinfo {year} {2022})}\BibitemShut
  {NoStop}%
\bibitem [{\citenamefont {Haque}\ \emph {et~al.}(2019)\citenamefont {Haque},
  \citenamefont {Hossain},\ and\ \citenamefont {Stampfl}}]{haque2019first}%
  \BibitemOpen
  \bibfield  {author} {\bibinfo {author} {\bibfnamefont {E.}~\bibnamefont
  {Haque}}, \bibinfo {author} {\bibfnamefont {M.~A.}\ \bibnamefont {Hossain}},\
  and\ \bibinfo {author} {\bibfnamefont {C.}~\bibnamefont {Stampfl}},\
  }\href@noop {} {\bibfield  {journal} {\bibinfo  {journal} {Physical Chemistry
  Chemical Physics}\ }\textbf {\bibinfo {volume} {21}},\ \bibinfo {pages}
  {8767} (\bibinfo {year} {2019})}\BibitemShut {NoStop}%
\bibitem [{\citenamefont {Novko}\ \emph {et~al.}(2020)\citenamefont {Novko},
  \citenamefont {Caruso}, \citenamefont {Draxl},\ and\ \citenamefont
  {Cappelluti}}]{novko2020ultrafast}%
  \BibitemOpen
  \bibfield  {author} {\bibinfo {author} {\bibfnamefont {D.}~\bibnamefont
  {Novko}}, \bibinfo {author} {\bibfnamefont {F.}~\bibnamefont {Caruso}},
  \bibinfo {author} {\bibfnamefont {C.}~\bibnamefont {Draxl}},\ and\ \bibinfo
  {author} {\bibfnamefont {E.}~\bibnamefont {Cappelluti}},\ }\href@noop {}
  {\bibfield  {journal} {\bibinfo  {journal} {Physical Review Letters}\
  }\textbf {\bibinfo {volume} {124}},\ \bibinfo {pages} {077001} (\bibinfo
  {year} {2020})}\BibitemShut {NoStop}%
\bibitem [{\citenamefont {Liu}\ \emph {et~al.}(2001)\citenamefont {Liu},
  \citenamefont {Mazin},\ and\ \citenamefont {Kortus}}]{liu2001beyond}%
  \BibitemOpen
  \bibfield  {author} {\bibinfo {author} {\bibfnamefont {A.~Y.}\ \bibnamefont
  {Liu}}, \bibinfo {author} {\bibfnamefont {I.}~\bibnamefont {Mazin}},\ and\
  \bibinfo {author} {\bibfnamefont {J.}~\bibnamefont {Kortus}},\ }\href@noop {}
  {\bibfield  {journal} {\bibinfo  {journal} {Physical Review Letters}\
  }\textbf {\bibinfo {volume} {87}},\ \bibinfo {pages} {087005} (\bibinfo
  {year} {2001})}\BibitemShut {NoStop}%
\bibitem [{\citenamefont {Quan}\ and\ \citenamefont
  {Pickett}(2020)}]{quan2020li}%
  \BibitemOpen
  \bibfield  {author} {\bibinfo {author} {\bibfnamefont {Y.}~\bibnamefont
  {Quan}}\ and\ \bibinfo {author} {\bibfnamefont {W.~E.}\ \bibnamefont
  {Pickett}},\ }\href@noop {} {\bibfield  {journal} {\bibinfo  {journal}
  {Physical Review B}\ }\textbf {\bibinfo {volume} {102}},\ \bibinfo {pages}
  {144504} (\bibinfo {year} {2020})}\BibitemShut {NoStop}%
\bibitem [{\citenamefont {Modak}\ \emph {et~al.}(2021)\citenamefont {Modak},
  \citenamefont {Verma},\ and\ \citenamefont {Mishra}}]{modak2021prediction}%
  \BibitemOpen
  \bibfield  {author} {\bibinfo {author} {\bibfnamefont {P.}~\bibnamefont
  {Modak}}, \bibinfo {author} {\bibfnamefont {A.~K.}\ \bibnamefont {Verma}},\
  and\ \bibinfo {author} {\bibfnamefont {A.~K.}\ \bibnamefont {Mishra}},\
  }\href@noop {} {\bibfield  {journal} {\bibinfo  {journal} {Physical Review
  B}\ }\textbf {\bibinfo {volume} {104}},\ \bibinfo {pages} {054504} (\bibinfo
  {year} {2021})}\BibitemShut {NoStop}%
\bibitem [{\citenamefont {Gao}\ \emph {et~al.}(2015)\citenamefont {Gao},
  \citenamefont {Lu},\ and\ \citenamefont {Xiang}}]{gao2015prediction}%
  \BibitemOpen
  \bibfield  {author} {\bibinfo {author} {\bibfnamefont {M.}~\bibnamefont
  {Gao}}, \bibinfo {author} {\bibfnamefont {Z.-Y.}\ \bibnamefont {Lu}},\ and\
  \bibinfo {author} {\bibfnamefont {T.}~\bibnamefont {Xiang}},\ }\href@noop {}
  {\bibfield  {journal} {\bibinfo  {journal} {Physical Review B}\ }\textbf
  {\bibinfo {volume} {91}},\ \bibinfo {pages} {045132} (\bibinfo {year}
  {2015})}\BibitemShut {NoStop}%
\bibitem [{\citenamefont {Di~Cataldo}\ \emph {et~al.}(2022)\citenamefont
  {Di~Cataldo}, \citenamefont {Qulaghasi}, \citenamefont {Bachelet},\ and\
  \citenamefont {Boeri}}]{di2022high}%
  \BibitemOpen
  \bibfield  {author} {\bibinfo {author} {\bibfnamefont {S.}~\bibnamefont
  {Di~Cataldo}}, \bibinfo {author} {\bibfnamefont {S.}~\bibnamefont
  {Qulaghasi}}, \bibinfo {author} {\bibfnamefont {G.~B.}\ \bibnamefont
  {Bachelet}},\ and\ \bibinfo {author} {\bibfnamefont {L.}~\bibnamefont
  {Boeri}},\ }\href@noop {} {\bibfield  {journal} {\bibinfo  {journal}
  {Physical Review B}\ }\textbf {\bibinfo {volume} {105}},\ \bibinfo {pages}
  {064516} (\bibinfo {year} {2022})}\BibitemShut {NoStop}%
\bibitem [{\citenamefont {Choi}\ \emph {et~al.}(2009)\citenamefont {Choi},
  \citenamefont {Louie},\ and\ \citenamefont {Cohen}}]{choi2009prediction}%
  \BibitemOpen
  \bibfield  {author} {\bibinfo {author} {\bibfnamefont {H.~J.}\ \bibnamefont
  {Choi}}, \bibinfo {author} {\bibfnamefont {S.~G.}\ \bibnamefont {Louie}},\
  and\ \bibinfo {author} {\bibfnamefont {M.~L.}\ \bibnamefont {Cohen}},\
  }\href@noop {} {\bibfield  {journal} {\bibinfo  {journal} {Physical Review
  B}\ }\textbf {\bibinfo {volume} {80}},\ \bibinfo {pages} {064503} (\bibinfo
  {year} {2009})}\BibitemShut {NoStop}%
\bibitem [{\citenamefont {Bekaert}\ \emph {et~al.}(2019)\citenamefont
  {Bekaert}, \citenamefont {Petrov}, \citenamefont {Aperis}, \citenamefont
  {Oppeneer},\ and\ \citenamefont {Milo{\v{s}}evi{\'c}}}]{bekaert2019hydrogen}%
  \BibitemOpen
  \bibfield  {author} {\bibinfo {author} {\bibfnamefont {J.}~\bibnamefont
  {Bekaert}}, \bibinfo {author} {\bibfnamefont {M.}~\bibnamefont {Petrov}},
  \bibinfo {author} {\bibfnamefont {A.}~\bibnamefont {Aperis}}, \bibinfo
  {author} {\bibfnamefont {P.~M.}\ \bibnamefont {Oppeneer}},\ and\ \bibinfo
  {author} {\bibfnamefont {M.}~\bibnamefont {Milo{\v{s}}evi{\'c}}},\
  }\href@noop {} {\bibfield  {journal} {\bibinfo  {journal} {Physical review
  letters}\ }\textbf {\bibinfo {volume} {123}},\ \bibinfo {pages} {077001}
  (\bibinfo {year} {2019})}\BibitemShut {NoStop}%
\bibitem [{\citenamefont {Bekaert}\ \emph {et~al.}(2017)\citenamefont
  {Bekaert}, \citenamefont {Aperis}, \citenamefont {Partoens}, \citenamefont
  {Oppeneer},\ and\ \citenamefont
  {Milo{\v{s}}evi{\'c}}}]{bekaert2017evolution}%
  \BibitemOpen
  \bibfield  {author} {\bibinfo {author} {\bibfnamefont {J.}~\bibnamefont
  {Bekaert}}, \bibinfo {author} {\bibfnamefont {A.}~\bibnamefont {Aperis}},
  \bibinfo {author} {\bibfnamefont {B.}~\bibnamefont {Partoens}}, \bibinfo
  {author} {\bibfnamefont {P.~M.}\ \bibnamefont {Oppeneer}},\ and\ \bibinfo
  {author} {\bibfnamefont {M.}~\bibnamefont {Milo{\v{s}}evi{\'c}}},\
  }\href@noop {} {\bibfield  {journal} {\bibinfo  {journal} {Physical Review
  B}\ }\textbf {\bibinfo {volume} {96}},\ \bibinfo {pages} {094510} (\bibinfo
  {year} {2017})}\BibitemShut {NoStop}%
\bibitem [{\citenamefont {Bazhirov}\ \emph {et~al.}(2014)\citenamefont
  {Bazhirov}, \citenamefont {Sakai}, \citenamefont {Saito},\ and\ \citenamefont
  {Cohen}}]{bazhirov2014electron}%
  \BibitemOpen
  \bibfield  {author} {\bibinfo {author} {\bibfnamefont {T.}~\bibnamefont
  {Bazhirov}}, \bibinfo {author} {\bibfnamefont {Y.}~\bibnamefont {Sakai}},
  \bibinfo {author} {\bibfnamefont {S.}~\bibnamefont {Saito}},\ and\ \bibinfo
  {author} {\bibfnamefont {M.~L.}\ \bibnamefont {Cohen}},\ }\href@noop {}
  {\bibfield  {journal} {\bibinfo  {journal} {Physical Review B}\ }\textbf
  {\bibinfo {volume} {89}},\ \bibinfo {pages} {045136} (\bibinfo {year}
  {2014})}\BibitemShut {NoStop}%
\bibitem [{\citenamefont {Rosner}\ \emph {et~al.}(2002)\citenamefont {Rosner},
  \citenamefont {Kitaigorodsky},\ and\ \citenamefont
  {Pickett}}]{rosner2002prediction}%
  \BibitemOpen
  \bibfield  {author} {\bibinfo {author} {\bibfnamefont {H.}~\bibnamefont
  {Rosner}}, \bibinfo {author} {\bibfnamefont {A.}~\bibnamefont
  {Kitaigorodsky}},\ and\ \bibinfo {author} {\bibfnamefont {W.}~\bibnamefont
  {Pickett}},\ }\href@noop {} {\bibfield  {journal} {\bibinfo  {journal}
  {Physical review letters}\ }\textbf {\bibinfo {volume} {88}},\ \bibinfo
  {pages} {127001} (\bibinfo {year} {2002})}\BibitemShut {NoStop}%
\bibitem [{\citenamefont {Fogg}\ \emph {et~al.}(2003)\citenamefont {Fogg},
  \citenamefont {Chalker}, \citenamefont {Claridge}, \citenamefont {Darling},\
  and\ \citenamefont {Rosseinsky}}]{fogg2003libc}%
  \BibitemOpen
  \bibfield  {author} {\bibinfo {author} {\bibfnamefont {A.}~\bibnamefont
  {Fogg}}, \bibinfo {author} {\bibfnamefont {P.}~\bibnamefont {Chalker}},
  \bibinfo {author} {\bibfnamefont {J.}~\bibnamefont {Claridge}}, \bibinfo
  {author} {\bibfnamefont {G.}~\bibnamefont {Darling}},\ and\ \bibinfo {author}
  {\bibfnamefont {M.}~\bibnamefont {Rosseinsky}},\ }\href@noop {} {\bibfield
  {journal} {\bibinfo  {journal} {Physical Review B}\ }\textbf {\bibinfo
  {volume} {67}},\ \bibinfo {pages} {245106} (\bibinfo {year}
  {2003})}\BibitemShut {NoStop}%
\bibitem [{\citenamefont {Haas}\ \emph {et~al.}(2019)\citenamefont {Haas},
  \citenamefont {Fischer}, \citenamefont {Hauf}, \citenamefont {Wieser},
  \citenamefont {Schmidt}, \citenamefont {Eickerling}, \citenamefont {Scheidt},
  \citenamefont {Schiffmann}, \citenamefont {Reckeweg}, \citenamefont {DiSalvo}
  \emph {et~al.}}]{haas2019color}%
  \BibitemOpen
  \bibfield  {author} {\bibinfo {author} {\bibfnamefont {C.~D.}\ \bibnamefont
  {Haas}}, \bibinfo {author} {\bibfnamefont {A.}~\bibnamefont {Fischer}},
  \bibinfo {author} {\bibfnamefont {C.}~\bibnamefont {Hauf}}, \bibinfo {author}
  {\bibfnamefont {C.}~\bibnamefont {Wieser}}, \bibinfo {author} {\bibfnamefont
  {A.~P.}\ \bibnamefont {Schmidt}}, \bibinfo {author} {\bibfnamefont
  {G.}~\bibnamefont {Eickerling}}, \bibinfo {author} {\bibfnamefont {E.-W.}\
  \bibnamefont {Scheidt}}, \bibinfo {author} {\bibfnamefont {J.~G.}\
  \bibnamefont {Schiffmann}}, \bibinfo {author} {\bibfnamefont
  {O.}~\bibnamefont {Reckeweg}}, \bibinfo {author} {\bibfnamefont {F.~J.}\
  \bibnamefont {DiSalvo}}, \emph {et~al.},\ }\href@noop {} {\bibfield
  {journal} {\bibinfo  {journal} {Angewandte Chemie}\ }\textbf {\bibinfo
  {volume} {131}},\ \bibinfo {pages} {2382} (\bibinfo {year}
  {2019})}\BibitemShut {NoStop}%
\bibitem [{\citenamefont {Hohenberg}\ and\ \citenamefont
  {Kohn}(1964)}]{DFT01_hohenberg1964inhomogeneous}%
  \BibitemOpen
  \bibfield  {author} {\bibinfo {author} {\bibfnamefont {P.}~\bibnamefont
  {Hohenberg}}\ and\ \bibinfo {author} {\bibfnamefont {W.}~\bibnamefont
  {Kohn}},\ }\href@noop {} {\bibfield  {journal} {\bibinfo  {journal} {Physical
  review}\ }\textbf {\bibinfo {volume} {136}},\ \bibinfo {pages} {B864}
  (\bibinfo {year} {1964})}\BibitemShut {NoStop}%
\bibitem [{\citenamefont {Kohn}\ and\ \citenamefont
  {Sham}(1965)}]{dft1965_kohn1965self}%
  \BibitemOpen
  \bibfield  {author} {\bibinfo {author} {\bibfnamefont {W.}~\bibnamefont
  {Kohn}}\ and\ \bibinfo {author} {\bibfnamefont {L.~J.}\ \bibnamefont
  {Sham}},\ }\href@noop {} {\bibfield  {journal} {\bibinfo  {journal} {Physical
  review}\ }\textbf {\bibinfo {volume} {140}},\ \bibinfo {pages} {A1133}
  (\bibinfo {year} {1965})}\BibitemShut {NoStop}%
\bibitem [{\citenamefont {Giannozzi}\ \emph {et~al.}(2020)\citenamefont
  {Giannozzi}, \citenamefont {Baseggio}, \citenamefont {Bonf{\`a}},
  \citenamefont {Brunato}, \citenamefont {Car}, \citenamefont {Carnimeo},
  \citenamefont {Cavazzoni}, \citenamefont {De~Gironcoli}, \citenamefont
  {Delugas}, \citenamefont {Ferrari~Ruffino} \emph
  {et~al.}}]{QUANTUMESPRESO03_giannozzi2020quantum}%
  \BibitemOpen
  \bibfield  {author} {\bibinfo {author} {\bibfnamefont {P.}~\bibnamefont
  {Giannozzi}}, \bibinfo {author} {\bibfnamefont {O.}~\bibnamefont {Baseggio}},
  \bibinfo {author} {\bibfnamefont {P.}~\bibnamefont {Bonf{\`a}}}, \bibinfo
  {author} {\bibfnamefont {D.}~\bibnamefont {Brunato}}, \bibinfo {author}
  {\bibfnamefont {R.}~\bibnamefont {Car}}, \bibinfo {author} {\bibfnamefont
  {I.}~\bibnamefont {Carnimeo}}, \bibinfo {author} {\bibfnamefont
  {C.}~\bibnamefont {Cavazzoni}}, \bibinfo {author} {\bibfnamefont
  {S.}~\bibnamefont {De~Gironcoli}}, \bibinfo {author} {\bibfnamefont
  {P.}~\bibnamefont {Delugas}}, \bibinfo {author} {\bibfnamefont
  {F.}~\bibnamefont {Ferrari~Ruffino}}, \emph {et~al.},\ }\href@noop {}
  {\bibfield  {journal} {\bibinfo  {journal} {The Journal of chemical physics}\
  }\textbf {\bibinfo {volume} {152}},\ \bibinfo {pages} {154105} (\bibinfo
  {year} {2020})}\BibitemShut {NoStop}%
\bibitem [{\citenamefont {Giannozzi}\ \emph {et~al.}(2017)\citenamefont
  {Giannozzi}, \citenamefont {Andreussi}, \citenamefont {Brumme}, \citenamefont
  {Bunau}, \citenamefont {Nardelli}, \citenamefont {Calandra}, \citenamefont
  {Car}, \citenamefont {Cavazzoni}, \citenamefont {Ceresoli}, \citenamefont
  {Cococcioni} \emph {et~al.}}]{QUANTUMESPRESSO02_giannozzi2017advanced}%
  \BibitemOpen
  \bibfield  {author} {\bibinfo {author} {\bibfnamefont {P.}~\bibnamefont
  {Giannozzi}}, \bibinfo {author} {\bibfnamefont {O.}~\bibnamefont
  {Andreussi}}, \bibinfo {author} {\bibfnamefont {T.}~\bibnamefont {Brumme}},
  \bibinfo {author} {\bibfnamefont {O.}~\bibnamefont {Bunau}}, \bibinfo
  {author} {\bibfnamefont {M.~B.}\ \bibnamefont {Nardelli}}, \bibinfo {author}
  {\bibfnamefont {M.}~\bibnamefont {Calandra}}, \bibinfo {author}
  {\bibfnamefont {R.}~\bibnamefont {Car}}, \bibinfo {author} {\bibfnamefont
  {C.}~\bibnamefont {Cavazzoni}}, \bibinfo {author} {\bibfnamefont
  {D.}~\bibnamefont {Ceresoli}}, \bibinfo {author} {\bibfnamefont
  {M.}~\bibnamefont {Cococcioni}}, \emph {et~al.},\ }\href@noop {} {\bibfield
  {journal} {\bibinfo  {journal} {Journal of physics: Condensed matter}\
  }\textbf {\bibinfo {volume} {29}},\ \bibinfo {pages} {465901} (\bibinfo
  {year} {2017})}\BibitemShut {NoStop}%
\bibitem [{\citenamefont {Yu}\ and\ \citenamefont
  {Trinkle}(2011)}]{yu2011accurate}%
  \BibitemOpen
  \bibfield  {author} {\bibinfo {author} {\bibfnamefont {M.}~\bibnamefont
  {Yu}}\ and\ \bibinfo {author} {\bibfnamefont {D.~R.}\ \bibnamefont
  {Trinkle}},\ }\href@noop {} {\bibfield  {journal} {\bibinfo  {journal} {The
  Journal of chemical physics}\ }\textbf {\bibinfo {volume} {134}},\ \bibinfo
  {pages} {064111} (\bibinfo {year} {2011})}\BibitemShut {NoStop}%
\bibitem [{\citenamefont {Perdew}\ \emph {et~al.}(1996)\citenamefont {Perdew},
  \citenamefont {Burke},\ and\ \citenamefont
  {Ernzerhof}}]{PBE_perdew1996generalized}%
  \BibitemOpen
  \bibfield  {author} {\bibinfo {author} {\bibfnamefont {J.~P.}\ \bibnamefont
  {Perdew}}, \bibinfo {author} {\bibfnamefont {K.}~\bibnamefont {Burke}},\ and\
  \bibinfo {author} {\bibfnamefont {M.}~\bibnamefont {Ernzerhof}},\ }\href@noop
  {} {\bibfield  {journal} {\bibinfo  {journal} {Physical review letters}\
  }\textbf {\bibinfo {volume} {77}},\ \bibinfo {pages} {3865} (\bibinfo {year}
  {1996})}\BibitemShut {NoStop}%
\bibitem [{\citenamefont {Prandini}\ \emph {et~al.}(2018)\citenamefont
  {Prandini}, \citenamefont {Marrazzo}, \citenamefont {Castelli}, \citenamefont
  {Mounet},\ and\ \citenamefont {Marzari}}]{SSSP01_prandini2018precision}%
  \BibitemOpen
  \bibfield  {author} {\bibinfo {author} {\bibfnamefont {G.}~\bibnamefont
  {Prandini}}, \bibinfo {author} {\bibfnamefont {A.}~\bibnamefont {Marrazzo}},
  \bibinfo {author} {\bibfnamefont {I.~E.}\ \bibnamefont {Castelli}}, \bibinfo
  {author} {\bibfnamefont {N.}~\bibnamefont {Mounet}},\ and\ \bibinfo {author}
  {\bibfnamefont {N.}~\bibnamefont {Marzari}},\ }\href@noop {} {\bibfield
  {journal} {\bibinfo  {journal} {npj Computational Materials}\ }\textbf
  {\bibinfo {volume} {4}},\ \bibinfo {pages} {1} (\bibinfo {year}
  {2018})}\BibitemShut {NoStop}%
\bibitem [{\citenamefont {Monkhorst}\ and\ \citenamefont
  {Pack}(1976)}]{KPOINTS_monkhorst1976special}%
  \BibitemOpen
  \bibfield  {author} {\bibinfo {author} {\bibfnamefont {H.~J.}\ \bibnamefont
  {Monkhorst}}\ and\ \bibinfo {author} {\bibfnamefont {J.~D.}\ \bibnamefont
  {Pack}},\ }\href@noop {} {\bibfield  {journal} {\bibinfo  {journal} {Physical
  review B}\ }\textbf {\bibinfo {volume} {13}},\ \bibinfo {pages} {5188}
  (\bibinfo {year} {1976})}\BibitemShut {NoStop}%
\bibitem [{\citenamefont {Baroni}\ \emph {et~al.}(2001)\citenamefont {Baroni},
  \citenamefont {De~Gironcoli}, \citenamefont {Dal~Corso},\ and\ \citenamefont
  {Giannozzi}}]{baroni2001phonons}%
  \BibitemOpen
  \bibfield  {author} {\bibinfo {author} {\bibfnamefont {S.}~\bibnamefont
  {Baroni}}, \bibinfo {author} {\bibfnamefont {S.}~\bibnamefont
  {De~Gironcoli}}, \bibinfo {author} {\bibfnamefont {A.}~\bibnamefont
  {Dal~Corso}},\ and\ \bibinfo {author} {\bibfnamefont {P.}~\bibnamefont
  {Giannozzi}},\ }\href@noop {} {\bibfield  {journal} {\bibinfo  {journal}
  {Reviews of modern Physics}\ }\textbf {\bibinfo {volume} {73}},\ \bibinfo
  {pages} {515} (\bibinfo {year} {2001})}\BibitemShut {NoStop}%
\bibitem [{\citenamefont {Eliashberg}(1960)}]{eliashberg1960interactions}%
  \BibitemOpen
  \bibfield  {author} {\bibinfo {author} {\bibfnamefont {G.}~\bibnamefont
  {Eliashberg}},\ }\href@noop {} {\bibfield  {journal} {\bibinfo  {journal}
  {Sov. Phys. JETP}\ }\textbf {\bibinfo {volume} {11}},\ \bibinfo {pages} {696}
  (\bibinfo {year} {1960})}\BibitemShut {NoStop}%
\bibitem [{\citenamefont {Wierzbowska}\ \emph {et~al.}(2005)\citenamefont
  {Wierzbowska}, \citenamefont {de~Gironcoli},\ and\ \citenamefont
  {Giannozzi}}]{wierzbowska2005origins}%
  \BibitemOpen
  \bibfield  {author} {\bibinfo {author} {\bibfnamefont {M.}~\bibnamefont
  {Wierzbowska}}, \bibinfo {author} {\bibfnamefont {S.}~\bibnamefont
  {de~Gironcoli}},\ and\ \bibinfo {author} {\bibfnamefont {P.}~\bibnamefont
  {Giannozzi}},\ }\href@noop {} {\bibfield  {journal} {\bibinfo  {journal}
  {arXiv preprint cond-mat/0504077}\ } (\bibinfo {year} {2005})}\BibitemShut
  {NoStop}%
\bibitem [{\citenamefont {Thonhauser}\ \emph {et~al.}(2015)\citenamefont
  {Thonhauser}, \citenamefont {Zuluaga}, \citenamefont {Arter}, \citenamefont
  {Berland}, \citenamefont {Schr{\"o}der},\ and\ \citenamefont
  {Hyldgaard}}]{thonhauser2015spin}%
  \BibitemOpen
  \bibfield  {author} {\bibinfo {author} {\bibfnamefont {T.}~\bibnamefont
  {Thonhauser}}, \bibinfo {author} {\bibfnamefont {S.}~\bibnamefont {Zuluaga}},
  \bibinfo {author} {\bibfnamefont {C.}~\bibnamefont {Arter}}, \bibinfo
  {author} {\bibfnamefont {K.}~\bibnamefont {Berland}}, \bibinfo {author}
  {\bibfnamefont {E.}~\bibnamefont {Schr{\"o}der}},\ and\ \bibinfo {author}
  {\bibfnamefont {P.}~\bibnamefont {Hyldgaard}},\ }\href@noop {} {\bibfield
  {journal} {\bibinfo  {journal} {Physical review letters}\ }\textbf {\bibinfo
  {volume} {115}},\ \bibinfo {pages} {136402} (\bibinfo {year}
  {2015})}\BibitemShut {NoStop}%
\bibitem [{\citenamefont {Yamashita}\ \emph {et~al.}(2022)\citenamefont
  {Yamashita}, \citenamefont {Kino}, \citenamefont {Tsuda}, \citenamefont
  {Miyake},\ and\ \citenamefont {Oguchi}}]{yamashita2022hybrid}%
  \BibitemOpen
  \bibfield  {author} {\bibinfo {author} {\bibfnamefont {T.}~\bibnamefont
  {Yamashita}}, \bibinfo {author} {\bibfnamefont {H.}~\bibnamefont {Kino}},
  \bibinfo {author} {\bibfnamefont {K.}~\bibnamefont {Tsuda}}, \bibinfo
  {author} {\bibfnamefont {T.}~\bibnamefont {Miyake}},\ and\ \bibinfo {author}
  {\bibfnamefont {T.}~\bibnamefont {Oguchi}},\ }\href@noop {} {\bibfield
  {journal} {\bibinfo  {journal} {Science and Technology of Advanced Materials:
  Methods}\ }\textbf {\bibinfo {volume} {2}},\ \bibinfo {pages} {67} (\bibinfo
  {year} {2022})}\BibitemShut {NoStop}%
\bibitem [{\citenamefont {Zhou}\ \emph {et~al.}(2014)\citenamefont {Zhou},
  \citenamefont {Dong}, \citenamefont {Oganov}, \citenamefont {Zhu},
  \citenamefont {Tian},\ and\ \citenamefont {Wang}}]{zhou2014semimetallic}%
  \BibitemOpen
  \bibfield  {author} {\bibinfo {author} {\bibfnamefont {X.-F.}\ \bibnamefont
  {Zhou}}, \bibinfo {author} {\bibfnamefont {X.}~\bibnamefont {Dong}}, \bibinfo
  {author} {\bibfnamefont {A.~R.}\ \bibnamefont {Oganov}}, \bibinfo {author}
  {\bibfnamefont {Q.}~\bibnamefont {Zhu}}, \bibinfo {author} {\bibfnamefont
  {Y.}~\bibnamefont {Tian}},\ and\ \bibinfo {author} {\bibfnamefont {H.-T.}\
  \bibnamefont {Wang}},\ }\href@noop {} {\bibfield  {journal} {\bibinfo
  {journal} {Physical Review Letters}\ }\textbf {\bibinfo {volume} {112}},\
  \bibinfo {pages} {085502} (\bibinfo {year} {2014})}\BibitemShut {NoStop}%
\bibitem [{\citenamefont {An}\ and\ \citenamefont
  {Pickett}(2001)}]{an2001superconductivity}%
  \BibitemOpen
  \bibfield  {author} {\bibinfo {author} {\bibfnamefont {J.}~\bibnamefont
  {An}}\ and\ \bibinfo {author} {\bibfnamefont {W.}~\bibnamefont {Pickett}},\
  }\href@noop {} {\bibfield  {journal} {\bibinfo  {journal} {Physical Review
  Letters}\ }\textbf {\bibinfo {volume} {86}},\ \bibinfo {pages} {4366}
  (\bibinfo {year} {2001})}\BibitemShut {NoStop}%
\bibitem [{\citenamefont {Kurmaev}\ \emph {et~al.}(2002)\citenamefont
  {Kurmaev}, \citenamefont {Lyakhovskaya}, \citenamefont {Kortus},
  \citenamefont {Moewes}, \citenamefont {Miyata}, \citenamefont {Demeter},
  \citenamefont {Neumann}, \citenamefont {Yanagihara}, \citenamefont
  {Watanabe}, \citenamefont {Muranaka} \emph {et~al.}}]{kurmaev2002electronic}%
  \BibitemOpen
  \bibfield  {author} {\bibinfo {author} {\bibfnamefont {E.}~\bibnamefont
  {Kurmaev}}, \bibinfo {author} {\bibfnamefont {I.}~\bibnamefont
  {Lyakhovskaya}}, \bibinfo {author} {\bibfnamefont {J.}~\bibnamefont
  {Kortus}}, \bibinfo {author} {\bibfnamefont {A.}~\bibnamefont {Moewes}},
  \bibinfo {author} {\bibfnamefont {N.}~\bibnamefont {Miyata}}, \bibinfo
  {author} {\bibfnamefont {M.}~\bibnamefont {Demeter}}, \bibinfo {author}
  {\bibfnamefont {M.}~\bibnamefont {Neumann}}, \bibinfo {author} {\bibfnamefont
  {M.}~\bibnamefont {Yanagihara}}, \bibinfo {author} {\bibfnamefont
  {M.}~\bibnamefont {Watanabe}}, \bibinfo {author} {\bibfnamefont
  {T.}~\bibnamefont {Muranaka}}, \emph {et~al.},\ }\href@noop {} {\bibfield
  {journal} {\bibinfo  {journal} {Physical Review B}\ }\textbf {\bibinfo
  {volume} {65}},\ \bibinfo {pages} {134509} (\bibinfo {year}
  {2002})}\BibitemShut {NoStop}%
\bibitem [{\citenamefont {Kim}\ \emph {et~al.}(2018)\citenamefont {Kim},
  \citenamefont {Ray}, \citenamefont {Bahr},\ and\ \citenamefont
  {Lordi}}]{kim2018electronic}%
  \BibitemOpen
  \bibfield  {author} {\bibinfo {author} {\bibfnamefont {C.-E.}\ \bibnamefont
  {Kim}}, \bibinfo {author} {\bibfnamefont {K.~G.}\ \bibnamefont {Ray}},
  \bibinfo {author} {\bibfnamefont {D.~F.}\ \bibnamefont {Bahr}},\ and\
  \bibinfo {author} {\bibfnamefont {V.}~\bibnamefont {Lordi}},\ }\href@noop {}
  {\bibfield  {journal} {\bibinfo  {journal} {Physical Review B}\ }\textbf
  {\bibinfo {volume} {97}},\ \bibinfo {pages} {195416} (\bibinfo {year}
  {2018})}\BibitemShut {NoStop}%
\bibitem [{\citenamefont {Ravindran}\ \emph {et~al.}(2001)\citenamefont
  {Ravindran}, \citenamefont {Vajeeston}, \citenamefont {Vidya}, \citenamefont
  {Kjekshus},\ and\ \citenamefont {Fjellv{\aa}g}}]{ravindran2001detailed}%
  \BibitemOpen
  \bibfield  {author} {\bibinfo {author} {\bibfnamefont {P.}~\bibnamefont
  {Ravindran}}, \bibinfo {author} {\bibfnamefont {P.}~\bibnamefont
  {Vajeeston}}, \bibinfo {author} {\bibfnamefont {R.}~\bibnamefont {Vidya}},
  \bibinfo {author} {\bibfnamefont {A.}~\bibnamefont {Kjekshus}},\ and\
  \bibinfo {author} {\bibfnamefont {H.}~\bibnamefont {Fjellv{\aa}g}},\
  }\href@noop {} {\bibfield  {journal} {\bibinfo  {journal} {Physical Review
  B}\ }\textbf {\bibinfo {volume} {64}},\ \bibinfo {pages} {224509} (\bibinfo
  {year} {2001})}\BibitemShut {NoStop}%
\bibitem [{\citenamefont {Yildirim}\ \emph {et~al.}(2001)\citenamefont
  {Yildirim}, \citenamefont {G{\"u}lseren}, \citenamefont {Lynn}, \citenamefont
  {Brown}, \citenamefont {Udovic}, \citenamefont {Huang}, \citenamefont
  {Rogado}, \citenamefont {Regan}, \citenamefont {Hayward}, \citenamefont
  {Slusky} \emph {et~al.}}]{yildirim2001giant}%
  \BibitemOpen
  \bibfield  {author} {\bibinfo {author} {\bibfnamefont {T.}~\bibnamefont
  {Yildirim}}, \bibinfo {author} {\bibfnamefont {O.}~\bibnamefont
  {G{\"u}lseren}}, \bibinfo {author} {\bibfnamefont {J.}~\bibnamefont {Lynn}},
  \bibinfo {author} {\bibfnamefont {C.}~\bibnamefont {Brown}}, \bibinfo
  {author} {\bibfnamefont {T.}~\bibnamefont {Udovic}}, \bibinfo {author}
  {\bibfnamefont {Q.}~\bibnamefont {Huang}}, \bibinfo {author} {\bibfnamefont
  {N.}~\bibnamefont {Rogado}}, \bibinfo {author} {\bibfnamefont
  {K.}~\bibnamefont {Regan}}, \bibinfo {author} {\bibfnamefont
  {M.}~\bibnamefont {Hayward}}, \bibinfo {author} {\bibfnamefont
  {J.}~\bibnamefont {Slusky}}, \emph {et~al.},\ }\href@noop {} {\bibfield
  {journal} {\bibinfo  {journal} {Physical review letters}\ }\textbf {\bibinfo
  {volume} {87}},\ \bibinfo {pages} {037001} (\bibinfo {year}
  {2001})}\BibitemShut {NoStop}%
\bibitem [{\citenamefont {Margine}\ and\ \citenamefont
  {Giustino}(2013)}]{margine2013anisotropic}%
  \BibitemOpen
  \bibfield  {author} {\bibinfo {author} {\bibfnamefont {E.~R.}\ \bibnamefont
  {Margine}}\ and\ \bibinfo {author} {\bibfnamefont {F.}~\bibnamefont
  {Giustino}},\ }\href@noop {} {\bibfield  {journal} {\bibinfo  {journal}
  {Physical Review B}\ }\textbf {\bibinfo {volume} {87}},\ \bibinfo {pages}
  {024505} (\bibinfo {year} {2013})}\BibitemShut {NoStop}%
\bibitem [{\citenamefont {An}\ \emph {et~al.}(2021)\citenamefont {An},
  \citenamefont {Li}, \citenamefont {Wang}, \citenamefont {Wang}, \citenamefont
  {Gong}, \citenamefont {Ma}, \citenamefont {Wang}, \citenamefont {Jiao},
  \citenamefont {Dong}, \citenamefont {Xu} \emph
  {et~al.}}]{an2021superconductivity}%
  \BibitemOpen
  \bibfield  {author} {\bibinfo {author} {\bibfnamefont {Y.}~\bibnamefont
  {An}}, \bibinfo {author} {\bibfnamefont {J.}~\bibnamefont {Li}}, \bibinfo
  {author} {\bibfnamefont {K.}~\bibnamefont {Wang}}, \bibinfo {author}
  {\bibfnamefont {G.}~\bibnamefont {Wang}}, \bibinfo {author} {\bibfnamefont
  {S.}~\bibnamefont {Gong}}, \bibinfo {author} {\bibfnamefont {C.}~\bibnamefont
  {Ma}}, \bibinfo {author} {\bibfnamefont {T.}~\bibnamefont {Wang}}, \bibinfo
  {author} {\bibfnamefont {Z.}~\bibnamefont {Jiao}}, \bibinfo {author}
  {\bibfnamefont {X.}~\bibnamefont {Dong}}, \bibinfo {author} {\bibfnamefont
  {G.}~\bibnamefont {Xu}}, \emph {et~al.},\ }\href@noop {} {\bibfield
  {journal} {\bibinfo  {journal} {Physical Review B}\ }\textbf {\bibinfo
  {volume} {104}},\ \bibinfo {pages} {134510} (\bibinfo {year}
  {2021})}\BibitemShut {NoStop}%
\bibitem [{\citenamefont {Kawamura}\ \emph {et~al.}(2014)\citenamefont
  {Kawamura}, \citenamefont {Gohda},\ and\ \citenamefont
  {Tsuneyuki}}]{kawamura2014improved}%
  \BibitemOpen
  \bibfield  {author} {\bibinfo {author} {\bibfnamefont {M.}~\bibnamefont
  {Kawamura}}, \bibinfo {author} {\bibfnamefont {Y.}~\bibnamefont {Gohda}},\
  and\ \bibinfo {author} {\bibfnamefont {S.}~\bibnamefont {Tsuneyuki}},\
  }\href@noop {} {\bibfield  {journal} {\bibinfo  {journal} {Physical Review
  B}\ }\textbf {\bibinfo {volume} {89}},\ \bibinfo {pages} {094515} (\bibinfo
  {year} {2014})}\BibitemShut {NoStop}%
\bibitem [{\citenamefont {Choi}\ \emph {et~al.}(2002)\citenamefont {Choi},
  \citenamefont {Roundy}, \citenamefont {Sun}, \citenamefont {Cohen},\ and\
  \citenamefont {Louie}}]{choi2002first}%
  \BibitemOpen
  \bibfield  {author} {\bibinfo {author} {\bibfnamefont {H.~J.}\ \bibnamefont
  {Choi}}, \bibinfo {author} {\bibfnamefont {D.}~\bibnamefont {Roundy}},
  \bibinfo {author} {\bibfnamefont {H.}~\bibnamefont {Sun}}, \bibinfo {author}
  {\bibfnamefont {M.~L.}\ \bibnamefont {Cohen}},\ and\ \bibinfo {author}
  {\bibfnamefont {S.~G.}\ \bibnamefont {Louie}},\ }\href@noop {} {\bibfield
  {journal} {\bibinfo  {journal} {Physical Review B}\ }\textbf {\bibinfo
  {volume} {66}},\ \bibinfo {pages} {020513} (\bibinfo {year}
  {2002})}\BibitemShut {NoStop}%
\bibitem [{\citenamefont {Bouquet}\ \emph {et~al.}(2001)\citenamefont
  {Bouquet}, \citenamefont {Fisher}, \citenamefont {Phillips}, \citenamefont
  {Hinks},\ and\ \citenamefont {Jorgensen}}]{bouquet2001specific}%
  \BibitemOpen
  \bibfield  {author} {\bibinfo {author} {\bibfnamefont {F.}~\bibnamefont
  {Bouquet}}, \bibinfo {author} {\bibfnamefont {R.}~\bibnamefont {Fisher}},
  \bibinfo {author} {\bibfnamefont {N.}~\bibnamefont {Phillips}}, \bibinfo
  {author} {\bibfnamefont {D.}~\bibnamefont {Hinks}},\ and\ \bibinfo {author}
  {\bibfnamefont {J.}~\bibnamefont {Jorgensen}},\ }\href@noop {} {\bibfield
  {journal} {\bibinfo  {journal} {Physical review letters}\ }\textbf {\bibinfo
  {volume} {87}},\ \bibinfo {pages} {047001} (\bibinfo {year}
  {2001})}\BibitemShut {NoStop}%
\end{thebibliography}
%

\end{document}